\documentclass[aps,prl,twocolumn,showpacs,preprintnumbers,
amsmath,amssymb]{revtex4}
\usepackage{epsfig}
\usepackage{graphicx}
\usepackage{dcolumn}
\usepackage{bm}

\begin{document}
\title{Magnetic field enhancement of superconductivity
in ultra-narrow wires}
\author{A. Rogachev, T.-C. Wei, D. Pekker, A.T. Bollinger, Paul M. Goldbart, and A. Bezryadin}
\affiliation{Department of Physics, University of Illinois at
Urbana-Champaign, Urbana, Illinois 61801-3080}
\date{\today}

\begin {abstract}
We study the effect of an applied magnetic field on sub-10nm wide
MoGe and Nb superconducting wires. We find that magnetic fields
can enhance the critical supercurrent at low temperatures, and
does so more strongly for narrower wires.  We conjecture that
magnetic moments are present, but their pair-breaking effect,
active at lower  magnetic fields, is suppressed by higher fields.
The corresponding microscopic theory, which we have developed,
quantitatively explains all experimental observations, and
suggests that magnetic moments have formed on the wire surfaces.
\end {abstract}


\maketitle

Magnetic fields have long been known to suppress superconductivity
through two main effects: first, by aligning the electron spins (i.e.,
the Zeeman effect) and second, by raising the kinetic energy of
electrons via Meissner screening currents (i.e., the orbital
effect)~\cite{TinkhamBook}.  However, some exceptions to this general
convention have been observed in nanoscale systems, where the field
can penetrate, essentially unattenuated, throughout the sample.  For
instance, a small applied magnetic field has been observed to cause
negative magneto-resistance (i.e., a decrease of resistance with
increasing field) in narrow superconducting strips~\cite{Dynes}.  More
strikingly, an applied magnetic field has been shown to strongly
enhance superconductivity in nanonwires: the so called anti-proximity
effect~\cite{mosesChan}.  We know of no commonly accepted theoretical
explanations for these effects in nanoscale systems.

Several (non-mutually exclusive) theoretical pictures have been
proposed for how magnetism or magnetic fields may enhance
superconductivity. First, the applied field may reduce the
charge-imbalance relaxation time associated with phase-slip
centers, thus resulting in negative magneto-resistance at high
currents and near \(T_\text{c}\)~\cite{clark}.  Second, the field
may enhance dissipative fluctuations, thus localizing the phase of
the superconductor and, thereby stabilizing
superconductivity~\cite{DHLee}; this  is thought to be relevant to
the anti-proximity effect~\cite{mosesChan}. Third, in disordered
superconductors having grain boundaries, negative
magneto-resistance may arise from interference between normal and
\(\pi\) junctions~\cite{Kivelson}. Finally, the pair-breaking
effect of magnetic moments may be quenched by either an applied
field~\cite{Kharitonov} or an exchange field~\cite{Canfield}, the
latter being relevant to magnetic superconductors.

In this Letter, we present results from experiments on
ultra-narrow, sub $10\,{\rm nm}$ wide MoGe and Nb homogeneous
superconducting wires that are nominally free of magnetic impurity
atoms.  We have found that at low temperatures magnetic fields can
enhance their critical currents by up to 30\%, reaching a maximum
at fields of ($2 -4\,{\rm T}$). This behavior is present both in
parallel and perpendicular field orientations, disappears at high
temperatures, and has the largest relative magnitude for the
thinnest wires. To explain this behavior we conjecture that
magnetic moments (due, e.g., to the surface oxide~\cite{Birge})
are present in the nanowires. Correspondingly, we have developed a
microscopic theory~\cite{Wei}, which shows that if such local
moments exist, a magnetic field can enhance the superconducting
critical current, $I_\text{c}$, and, in line with recent work by
Kharitonov and Feigel'man~\cite{Kharitonov}, raise $T_\text{c}$.
The essential physics involves the polarization of the magnetic
moments by the magnetic field~\cite{Glazman}, which quenches their
exchange-coupling with the electrons in Cooper pairs.  Our theory
is consistent with all our experimental observations, and also
suggests that in the present experiments the magnetic moments are
located on the surfaces of the wires.

\begin {table*}[t]
\begin {center}
\begin {tabular}{c|c c c c c c|c c|c c c c}
\hline \hline Sample & \ $t$ &\ $w$ & \ $L$ & \
$R_\text{N}$ (${\rm k}\Omega$)& \  $d$ & \ $I_\text{c}(0)$ (${\rm nA}$)& \ $T_\text{c}$ (${\rm K}$) &
\ $\xi(0)$ & \ $\tau_\text{B}$ (ps)& \ $d_\text{fit}$ & \ $T_{\text{c}0}$ (K)
& \ $I_\text{c}(0)/I_{\text{c},\text{fit}}(0)$ \\
\hline
MG1 ($\perp$)& $10$ & $21$ & $106$ & $2.14$ & $10.4$ & $1930$ & $3.8$ & $18$ & $3.6$ & $8.9$ & $5.0$ & $1.01$  \\
MG1 ($\parallel$)& $$ & $$ & $$ & $2.26$ & $10.0$ & $1760$ & $3.7$ & $19$ & $3.5$ & $8.9$ & $5.0$ & $0.92$ \\
MG2 ($\perp$)& $8$ & $17$ & $128$ & $3.24$ & $9.2$ & $1010$ & $3.6$ & $17$ & $2.4$ & $8.7$ & $5.6$ & $0.69$  \\
MG3 ($\perp$)& $7$ & $17.5$ & $156$ & $3.86$ & $9.4$ & $880$ & $2.9$ & $17$ & $3.4$ & $8.5$ & $4.4$ & $0.75$  \\
MG3 ($\parallel$)& $$ & $$ & $$ & $3.86$ & $9.4$ & $800$ & $2.9$ & $17$ & $3.1$ & $8.3$ & $4.4$ & $0.82$  \\
MG4 ($\parallel$)& $8$ & $12.5$ & $104$ & $4.84$ & $6.8$ & $63$ & $1.9$ & $39$ & $1.9$ & $9.1$ & $4.6$ & $0.22$  \\
Nb1 ($\perp$)& $7$ & $18$ & $120$ & $0.70$ & $8$ & $7170$ & $5.7$ & $8.1$ & $5.9$ & $6.4$ & $6.5$ & $0.89$  \\
Nb2 ($\perp$)& $4$ & $11$ & $110$ & $4.25$ & $3.1$ & $109$ & $1.5$ & $28.5$ & $4.9$ & $3.1$ & $2.5$ & $0.72$  \\
\hline \hline
\end {tabular}
\caption {\label {tableParameters} Summary of nanowire parameters
(all lengths are in \({\rm nm}\)).  The symbols ($\perp$) and
($\parallel$) indicate orientations of the magnetic field.  Wire
sample parameters: $t$--nominal thickness; $w$--width measured via
SEM; $L$--length; $R_N$--normal-state resistance; $d_R$--
diameter, estimated from $R_\text{N}$, $L$ and the resistivity of
MoGe ($170 \, \mu\Omega \, {\rm cm}$) and Nb ($30 \, \mu\Omega \,
{\rm cm}$)\cite{RogachevAPL}; $I_\text{c}(0)$--zero-field critical
current at \(0.3\, {\rm K}\). Parameters produced by the fitting
of $R$ vs.~$T$ curves at $B=0\,{\rm T}$ using TAPS theory:
$T_\text{c}$--critical temperature of the wire;
$\xi(0)$--superconducting coherence length in the wire. Parameters
used to fit our theory to $I_\text{c}(B)$ data
(Fig.~\ref{figIcB}): $\tau_\text{B}$--exchange-scattering time due
to local magnetic moments; $d_\text{fit}$--effective diameter of
wire, assuming circular cross-section; $T_{\text{c}0}$--critical
temperature of the wire without local moments;
$I_\text{c}(0)/I_{\text{c},\text{fit}}(0)$--rescaling factor.}
\end {center}
\end {table*}

To fabricate sub-10 nm wide wires we have used a
recently-developed molecular templating
technique~\cite{BezryadinDaiHopkins}. Our nanowires were made from
the superconducting amorphous alloy Mo$_{0.79}$Ge$_{0.21}$ or Nb,
deposited on to a free-standing carbon nanotube (or bundle of
tubes) suspended over a trench in a multilayered Si/SiO2/SiN
substrate. Combined fraction of Fe, Co and Ni was less than
$10^{-4}$ at.$\%$ in MoGe sputtering target and less than
$10^{-2}$ at.$\%$ in Nb target. The cross-section of the nanowire
is determined by the width of the templating nanotube and the
nominal thickness of material deposited. We exposed the MoGe wires
to the ambient atmosphere, which led to the oxidation of their
surfaces. This process reduces the width of the conducting core by
about $5\, {\rm nm}$~\cite{Bollinger}. Our Nb nanowires were
covered with protective Si layer~\cite{RogachevAPL}. The
parameters of the wires are given in Table~\ref{tableParameters}.

\begin{figure}[b]
\begin{center}
\epsfig{file=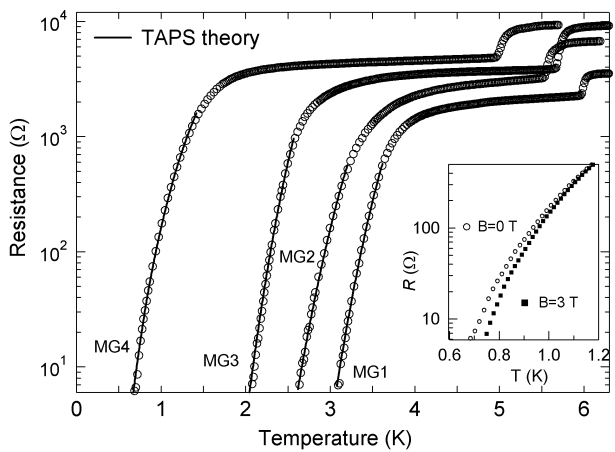, width=3.5 in}\label{fig:RT}
\vspace{-.5cm} \caption{\label{figRT} Temperature dependence of
the resistance of MoGe nanowires. For each sample, the solid line
indicates a fit to the TAPS theory~\cite{Sang}. Inset: \(R \)
vs.~\(T\) dependence of wire MG4 at  B=0 and $3\,{\rm T}$.}
\end{center}
\end{figure}

Electrical transport measurements were performed on the wires in a
\(^3\)He cryostat equipped with carefully filtered leads.  The
zero-bias resistance $R$ of the wires, measured in the
linear-response regime, is shown as a function of temperature $T$
in Fig.~\ref{figRT}. For each $R(T)$ curve, the higher-temperature
transition corresponds to the superconducting transition in the
film electrodes, which are connected in series with the wires. The
resistance measured immediately below the film transition is taken
as the normal-state resistance $R_\text{N}$ of the wire. Each
curve also shows a lower-temperature transition, corresponding to
the appearance of superconductivity in the wire itself. In
effectively one-dimensional superconductors, the resistance below
the critical temperature is never zero, owing to thermal
fluctuations that locally suppress the superconducting order
parameter, and thus allow current-dissipating phase slips
\cite{TinkhamBook}. To fit our resistance data we have used a
phenomenological formula that accounts for thermally activated
phase slips (TAPS) $R=R_\text{N}\exp[-\Delta
F/k_\text{B}T]$~\cite{Sang}, where $\Delta F$ is the free-energy
barrier for phase slips~\cite{LAMH}. The fitting parameters that
determine $\Delta F$ are $T_\text{c}$ and the zero-temperature
coherence length $\xi(0)$. We find excellent agreement with
experiment, which allows us to extract their values (see
Fig.~\ref{figRT} and Table~\ref{tableParameters}).

For thicker samples (MG1-MG3), increasing the magnetic field \(B\)
(not shown here) shifts the resistive transition of the wire to
progressively lower temperatures, in agreement with previously
observed behavior~\cite{RogachevPRL}. However, for the thinnest
sample (MG4), the $R(T)$ curve displays a more complex response to
the magnetic field: whereas at the highest fields ($B\approx 5
-9\, {\rm T}$) the aforementioned suppression of superconductivity
is observed, there is a regime of lower fields ($B\approx 0-3\,
{\rm T}$) for which the resistive transition of the wire shifts
oppositely, i.e., to higher temperatures with increasing \(B\), as
shown in the inset to Fig.~1.  This constitutes negative
magnetoresistance, which has also been observed in Pb
wires~\cite{Dynes}, and indicates that in this lower-field regime
the magnetic field enhances superconductivity.

\begin{figure}[t]
\begin{center}
\epsfig{file=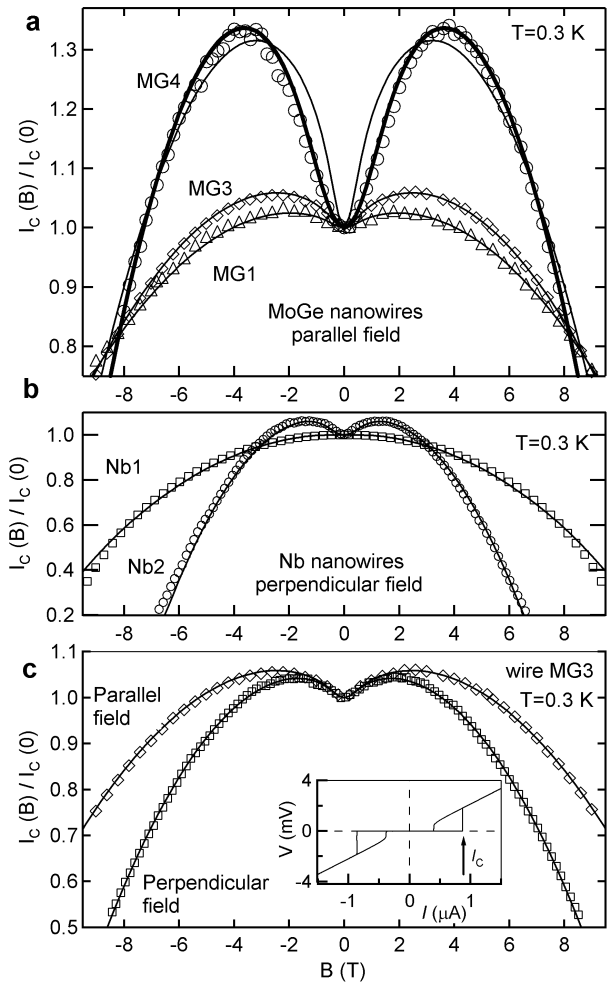, width=3.5 in} \vspace{-.5cm}
\caption{\label{figIcB}(a) Normalized critical current
vs.~magnetic field at $T=0.3\, {\rm K}$ for various MoGe wires
(parallel field). The thin solid lines are fits to our microscopic
theory (for the parameters see Table 1). The thick solid line
corresponds to a fit in which we allow variation of additional
parameters (the gyromagnetic ratio $g$ of local magnetic moments
($g=2$) and the average of the exchange coupling between electrons
and local magnetic moments divided by Fermi energy  $\langle
\widetilde{J}\rangle/E_F=-0.3$). (b) Normalized critical current
of Nb nanowires in perpendicular magnetic field. (c) Normalized
critical current of nanowire MG3 at $T=0.3\,{\rm K}$, measured in
parallel and perpendicular magnetic fields. Inset: A typical,
hysteretic voltage vs.~current curve. The transition from the
superconducting to the resistive state occurs via a single jump at
the indicated critical current.}
\end{center}
\end{figure}

In Fig.~\ref{figIcB}a we show the normalized critical currents of MoGe
wires of various diameters, measured in a parallel magnetic field at
$T=0.3\, {\rm K}$. Experimentally, $I_\text{c}$ is taken to be the
current at which the wire switches to the resistive state (see the inset
in Fig.~\ref{figIcB}c).
For all measured MoGe samples, $I_\text{c}$ displays
remarkable behavior, initially growing with increasing magnetic field
before reaching a maximum at $B\sim 2\,\, \text{to}\,\, 4\, {\rm T}$. The
relative magnitude of the enhancement of $I_\text{c}$ grows with the
reduction of the wire diameter; the largest enhancement (which occurs
for the thinnest wire, MG4) is about 30\% . Nanowires made of Nb
display the same tendency (see Fig.~2b). Whereas the thicker wire Nb1
shows the expected monotonic decrease of $I_\text{c}$, the critical
current anomaly is present in the much thinner wire Nb2.

To assess whether the effect is non-local in origin (e.g., is
associated with the pattern of supercurrent in the wire) we
applied both parallel and perpendicular magnetic fields to samples
MG3 and MG1 (always keeping the magnetic field parallel to the
electrode films). Between measurements in distinct
field-orientations the samples were removed from the cryostat,
rotated on the chip and rewired. After this procedure, the
zero-field critical current was found to decrease by about 10\%,
probably due to some shrinking of the cross-sectional area of the
wire via additional oxidation (see Table~\ref{tableParameters}).
The critical current for sample MG3, normalized by its value at
zero field, is shown in Fig.~\ref{figIcB}c. We found that the
initial rise is essentially the same for both field orientations.
This strongly suggests that the enhancement of $I_\text{c}$ is
local in origin.

To understand the anomalous enhancement of superconductivity by
magnetic fields in our nanowires, we have developed a theoretical
model (see Ref.~\cite{Wei}) that yields the dependence of
\(I_\text{c}\) on \(B\) and \(T\). We included the following
ingredients: (i)~local magnetic moments, which cause exchange
scattering of electrons (with zero-field exchange scattering time
given by $\tau_\text{B} = E_\text{F}/2 \pi
\langle\widetilde{J}^2\rangle x_\text{m}$, where \(x_\text{m}\) is
the fractional concentration of local moments, \(\widetilde{J}\)
is the exchange coupling, and \(E_\text{F}\) is the Fermi energy),
and thus lead to the breaking of Cooper pairs; (ii)~the vector
potential (associated with the applied magnetic field), which
scrambles the relative phases of the partners in a Cooper pair as
they move diffusively in the presence of impurity scattering
(viz., the orbital effect), which also suppresses
superconductivity; (iii)~the applied magnetic field, which
polarizes the local magnetic moments, and thus decreases the rate
of exchange-scattering, hence diminishing the contribution of
process (i) to de-pairing and thus enhancing superconductivity;
and (iv)~the Zeeman effect, associated with the applied field,
which splits the energy of the up and down spins of the Cooper
pair and thus tends to suppress superconductivity. (Note that
strong spin-orbit scattering tends to weaken de-pairing due to the
Zeeman effect.) These ingredients, which were also employed by
Kharitonov and Feigel'man in their work on critical temperatures,
embody the competing tendencies produced by the magnetic field:
de-pairing via the orbital and Zeeman effects, but also the
mollification of the de-pairing caused by local magnetic moments.

To obtain the critical current we first derived the semiclassical
Eilenberger-Usadel equations~\cite{Eilenberger} for the anomalous Green
function, taking into account terms that describe spin-orbit
scattering (with scattering time \(\tau_\text{SO}\)), local magnetic
moments, and the magnetic field~\cite{Wei}.  Then we seek the
current-carrying solution that maximizes the current, and identify it
as the critical de-paring current.  To fit the experimentally-measured
switching current we introduce the ratio of switching current to
de-pairing current $I_\text{c}(B)/I_{\text{c},\text{fit}}(B) (\le 1)$
as a fitting parameter, and assume that this ratio does not depend on
\(B\).

By carrying out this procedure for the case of spin-1/2 magnetic
impurities we have obtained numerical solutions for a wide range
of material parameters, temperatures and magnetic fields, and have
thus found three distinct regimes: a naturally expected one, in
which both  $I_\text{c}$ and $T_\text{c}$ simply decrease with
$B$; and two anomalous variants. The first gives non-monotonic
behavior for both $I_\text{c}$ and $T_\text{c}$, both first rising
and then falling with $B$. The second is even more striking:
although $T_\text{c}$ simply decreases with $B$, at low
temperatures  $I_\text{c}$ first rises and then falls.  Most of
our wires have behavior  in this last regime. To make a
quantitative comparison between our experiments and our theory, we
have performed fits to our data, allowing variations in the wire
diameter and the exchange scattering time.  For the remaining
parameters we have used following values: the $g$-factor $g=2$,
the spin-orbit scattering times $\tau_{\rm
so}=5.0\times10^{-14}\,{\rm s}$ for MoGe and $2.3\times10^{-12}$
for Nb~\cite{RogachevPRL}, and the diffusion constant for MoGe
$D=1\, {\rm cm}^{2}/{\rm s}$~\cite{Graybeal}. An important
consequence of our theory is that the initial behavior of the
$I_\text{c}(B)$ curves should not depend on the relative
orientation of the field and the wire.  This is because the
behavior of \(I_\text{c}\) at small \(B\) is dominated by
scattering from magnetic impurities, which is a local property and
thus insensitive to field orientation.  At larger fields the
orbital effect becomes important, and is larger for the
perpendicular field orientation. Hence, we expect that the
$I_\text{c}(B)$ curves should separate from one another, and that
the maximum in the parallel field orientation should occur at a
larger field than in the perpendicular orientation.
Figure~\ref{figIcB}c shows that our experimental data exhibit all
these properties. Further evidence in favor of our theoretical
picture comes from the fact that the fits to perpendicular- and
parallel-field data return essentially identical values for the
magnetic-impurity scattering time, which is proportional to the
impurity concentration, (Table~\ref{tableParameters}).

\begin{figure}[t]
\begin{center}
\epsfig{file=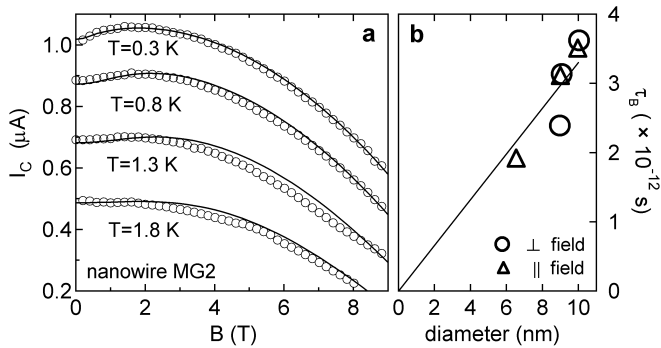, width=3.5 in} \vspace{-.0cm}
\label{figIcBT}\caption{(a) Critical current vs.~magnetic field
for various temperatures.  Solid lines are fits to the microscopic
theory. Only the $T=0.3\,{\rm K}$ curve was fitted. The same
microscopic parameters were used to generate curves at higher
temperatures.  The rescaling ratio
$I_\text{c}(0)/I_{\text{c},\text{fit}}(0)$ was adjusted at each
temperature. (b) Exchange scattering time vs.~wire diameter for
MoGe nanowires. The straight line is the linear fit.}
\end{center}
\end{figure}

At higher temperatures thermal fluctuations in the
moment-orientations make the quenching by the applied field less
effective and, hence, higher fields (at which the orbital effect
already becomes dominant) are required to quench the local
moments. Thus, the anomaly is expected to diminish. This is indeed
what we observe experimentally (~Fig.~3a): at our lowest
temperature (\(0.3\,{\rm K}\)) the anomaly is clearly observed;
but at temperatures higher than roughly $1.8\,{\rm K}$ the anomaly
is completely washed out.  This loss of the anomaly at higher
temperatures is consistent with the absence of any  observed
negative magneto-resistance for samples MG1-MG3, as their
resistances become too small to measure at the temperatures for
which the anomaly should appear.

Finally, in Fig.~3b we display $\tau_\text{B}$ as a function of
the wire diameter $d$.  Assuming that the magnitude of the
exchange integral $\widetilde{J}\approx0.2$ eV, we find the
magnetic-impurity fraction $x_\text{m}$ to be of order of 0.2 at.
\%. If the moments were distributed homogeneously throughout the
MoGe then $x_\text{m}$, and therefore $\tau_\text{B}$, would not
depend on the wire diameter.  Instead, data suggest that
$\tau_\text{B}$ depends linearly on $d$, consistent with the
magnetic moments being distributed over the surface of the wires.
This is also supported by the fact that our thick film Nb and MoGe
\cite{Bollinger}samples prepared under the same condition do not
reveal any change in $T_{c}$ compare to the published data
\cite{Graybeal}. Observation of anomalous behavior both in MoGe
and Nb nanowires suggests that such behavior is likely to occur
for nano-devices made from other superconducting materials, unless
suitable treatment is applied to avoid the formation of local
moments.

We thank M.Yu. Kharitonov and N.O. Birge for useful discussions.
This work was supported by DOE under Award No. DEFG02-91ER45439,
and by NSF under award No. EIA01-21568.


\end{document}